\documentclass{elsart}
\usepackage{epsfig}
\usepackage{colordvi}

\begin{document}

\def \ar {\texttt{AMBRE}{}}
\def \mb {\texttt{MB}{}}

\newcommand{\bea}{\begin{eqnarray}}
\newcommand{\eea}{\end{eqnarray}}
\newcommand{\bqa}{\begin{eqnarray}}
\newcommand{\eqa}{\end{eqnarray}}
\newcommand{\non}{\nonumber}
\newcommand{\nl}{\nonumber \\}
\newcommand{\ep}{\varepsilon}
\newcommand{\eps}{\varepsilon}
\newcommand{\RS}{\RawSienna}
\newcommand{\MG}{\Magenta}
\newcommand{\BL}{\Blue}

\newcommand{\bq}{ \begin {equation} }
\newcommand{\eq}{\end{equation}}
\newcommand{\be}{\begin{eqnarray}}
\newcommand{\ea}{\end{eqnarray}}

\newcommand{\nn}{\nonumber}

\def \Log {{\rm{ln}}}
\def \PolyLog {{\rm{Polylog}}}

\def \litwo {{\rm{Li_2}}}
\def \litr {{\rm{Li_3}}}
\def \lifo {{\rm{Li_4}}}
\def \ep   {\epsilon}
\def \litt {{\rm{S_{2,2}}}}
\def \liot {{\rm{S_{1,2}}}}

\texttt{
\begin{flushleft}
DESY 07-037 \hfill   
\\
HEPTOOLS 07-009
\\
SFB/CPP-07-14
\end{flushleft}}
\vspace*{1cm} 
\begin{frontmatter}
  \title {\bf \ar{} -- a Mathematica package for the construction of Mellin-Barnes representations
for Feynman integrals}
  \author{J. Gluza, K. Kajda}
    \address{Department of Field Theory and Particle Physics,
    Institute of Physics, \\
    University of Silesia, Uniwersytecka 4, PL-40-007 Katowice,
    Poland}
\author{T. Riemann}
\address{Deutsches Elektronen-Synchrotron, DESY, \\
   Platanenallee 6, 15738 Zeuthen, Germany}

\begin{abstract}
The Mathematica toolkit \ar{} derives  Mellin-Barnes (MB) representations
for Feynman integrals in $d=4-2\eps$ dimensions.
It may be applied for tadpoles as well as for multi-leg multi-loop 
scalar and tensor integrals. 
\ar{} uses a loop-by-loop approach and aims at lowest dimensions of the final MB  representations.
The present version of \ar{} works fine for planar Feynman diagrams.
The output may be further processed by the package {\ttfamily MB} for the determination of its singularity structure in $\eps$.
The \ar{} package contains various sample applications for Feynman integrals with up to six external particles and up 
to four loops.
\end{abstract}

\end{frontmatter}

\clearpage

\tableofcontents

\clearpage

\listoffigures

\clearpage
\section{\label{secintro}Introduction}
Recently, Mellin-Barnes (MB) representations of Feynman integrals have been used extensively in various phenomenological and theoretical studies of quantum field theory.
In many applications, sometimes in quite sophisticated ones  \cite{Smirnov:1999gc,Tausk:1999vh,Smirnov:2001cm},  the MB-integrals can be solved analytically.
One also may merge knowledge of some analytical solutions given by MB-integrals with other methods, 
e.g. the differential equations approach, as demonstrated in \cite{Czakon:2006hb}.
An introduction to the subject with many examples may be found in the monographies \cite{Smirnov:2004,Smirnov:2006ry}.
A systematic derivation and numerical evaluation of MB-representa\-tions for Feynman integrals with a (unpublished) Maple package was described in \cite{Anastasiou:2005cb}.
At the same time, the Mathematica program \mb{} for the automatized analytic continuation of MB-integrals was published in \cite{Czakon:2005rk}. 
With \ar, we deliver a Mathematica tool for the derivation of MB-integrals and their subsequent 
analytic continuation and numerical evaluation with \mb.

The article is organized as follows. 
In section \ref{sec-cons} we introduce the formulae used for the MB-representation of a general Feynman integral.
 The basic features of \ar{} are described in section 
\ref{sec:ambre}. 
One-loop examples are given in section \ref{sec:one-loop0}.
Section \ref{sec:beyond} 
describes the implementation of the loop-by-loop approach to multi-loop integrals.
Examples with tadpoles and on-shell diagrams as well as problems related to non-planar topologies  are discussed in sections \ref{sec:tadp}--\ref{sec:nonpl}.
A summary  follows in section \ref{sec:concl}.
In an appendix we list the Mathematica functions of \ar.

\section{\label{sec-cons}Construction of Mellin-Barnes representations}
The backbone of the procedure to build up MB-representations with \ar{} is the relation
\be
\frac{1}{(A+B)^{\nu}}
= \frac{B^{-\nu}}{2 \pi i\Gamma{(\nu)}}
\int\limits_{-i \infty}^{i \infty}
d \sigma {A^\sigma ~
B^{-\sigma}} ~ \Gamma{(-\sigma)}\Gamma{(\nu+\sigma)} ,
\label{mb}
\ea
where the integration contour separates the poles of the $\Gamma$-functions.

The object to be evaluated by \ar{} is an
$L$-loop Feynman integral\footnote{Often one uses the additional normalization $e^{\eps\gamma_E L}$; we leave this to the later evaluation with the package \mb{} \cite{Czakon:2005rk}.} in $d=4-2\eps$ dimensions with $N$ internal lines with momenta $q_i$ and masses  $m_i$,  and $E$
external legs with momenta $p_e$:
\bea\label{eq-bha}
G_L[T(k)]
=
\frac{1}{(i\pi^{d/2})^L} \int \frac{d^dk_1 \ldots d^dk_L~~T(k)}
     {(q_1^2-m_1^2)^{\nu_1} \ldots (q_i^2-m_i^2)^{\nu_j} \ldots
       (q_N^2-m_N^2)^{\nu_N}  }  .
\eea
The numerator $T(k)$ is a tensor in the integration variables:
\bea\label{eq-T}
T(k) &=& 1, k_l^{\mu}, k_l^{\mu}k_n^{\nu}, \ldots 
\eea 
The momenta of the denominator functions $d_i$ may be expressed by external and loop momenta:
\bea\label{di1} 
d_i &=& q_i^2-m_i^2 ~=~
\left(\sum_{l=1}^{L} \alpha_{il} k_l - P_i\right)^2 -m_i^2 
~=~ \left(\sum_{l=1}^{L} \alpha_{il} k_l -\sum_{e=1}^{E} \beta_{ie} p_e\right)^2 -m_i^2.
\eea
In the package \ar, in a first step the momentum integrals are replaced by Feynman parameter integrals:
\bea
\label{eq-scalar1}
G_L[T(k)]&=& 
\frac{(-1)^{N_{\nu}} \Gamma\left(N_{\nu}-\frac{d}{2}L\right)}
{\prod_{i=1}^{N}\Gamma(\nu_i)}
\int_0^1 \prod_{j=1}^N dx_j ~ x_j^{\nu_j-1}
\delta(1-\sum_{i=1}^N x_i)
\frac{U(x)^{N_{\nu}-d(L+1)/2}}{F(x)^{N_{\nu}-dL/2}}~P_L(T)
\nl
\eea
with
\bea\label{eq-Nnu}
N_{\nu} &=& \sum_{i=1}^{N} \nu_i.
\eea
The two functions $U$ and $F$ are 
characteristics of the topology of the Feynman integral.
One may derive them from
\bea
\label{bh-6}
{\mathcal N} &=& \sum_{i=1}^{N} x_i (q_i^2-m_i^2) ~\equiv~ kMk - 2kQ + J,
\eea
where $M_{ll'} = \sum_{i=1}^N \alpha_{il'}\alpha_{il} x_{i} $, and $Q_l = \sum_{i=1}^N \alpha_{il} P_i x_i$, and $J = \sum_{i=1}^N (P_i^2-m_i^2)x_i$;
namely:
\bea\label{eq:U} 
U{(x)} &=& \textrm{det}( M),
\\\label{eq:F} 
F({x}) &=& 
-\textrm{det}( M)~J + Q\tilde{M}Q.
\eea
The $U$ and $F$ as well as $\tilde{M} = \textrm{det}( M) ~ M^{-1} $ are polynomials in $x$,
and so are 
the numerator functions $P_L(T)$ in (\ref{eq-bha}) for scalar and vector integrals:
\bea
P_L(1) &=& 1,
\\
P_L(k_l^{\alpha}) &=&
\sum_{l'=1}^L {\tilde{M}}_{ll'}Q_{l'}^{\alpha}.
\eea
Tensors of higher degree 
depend additionally on the diagonalizing rotation $V$ for $ \mathcal N$,
\bea
{\mathcal N}_{diag} =  (\alpha_1,\ldots,\alpha_L) ~=~ (V^{-1})^+MV^{-1},
\eea
and become non-polynomial in $x$.
As an example, we quote here the case of an $L$-loop integral with a tensor of degree two:
\bea\label{eq:plt2} 
P_L(k_l^{\alpha}k_{l'}^{\beta})
&=&
\sum_{i=1}^L\Biggl[
 [{\tilde{M}}_{li}Q_i]^{\alpha}[{\tilde{M}}_{l'i}Q_{i}]^{\beta}
-
\frac{\Gamma\left(N_{\nu}-\frac{d}{2}L-1\right)}
{\Gamma\left(N_{\nu}-\frac{d}{2}L\right)} U F 
 \frac{(V_{li}^{-1})^+ (V_{l'i}^{-1})}   {\alpha_i}~\frac{g^{\alpha\beta}}{2} 
\Biggr].
\eea
The formulae simplify considerably for one-loop integrals:
\begin{eqnarray}
 \label{u1loop} 
U &=& M ~=~ {\tilde{M}} ~=~ \textrm{det}(M) ~=~ V ~=~ \sum_i^N x_i ~=~ 1,
\\ \label{f1loop} 
F &=& -U J+ Q^2 ~=~ \sum_{i,j}^N [P_iP_j-P_i^2+m_i^2]x_ix_j ~\equiv~  \sum_{i\leq j}^N f_{ij}x_ix_j .
\end{eqnarray} 
Then, the tensor factors  $P_1(T)$ in (\ref{eq-scalar1}) will become:
\bea
\label{p1loop}
P_1(1) &=& 1,
\\\label{eq:p1k} 
P_1(k^\alpha) &=& \sum_{i=1}^N x_i P_i^\alpha,
\\\label{eq:p1tens2} 
P_1(k^{\alpha}k^{\beta}) &=& \sum_{i=1}^N x_i P_i^{\alpha}\sum_{j=1}^N x_j P_j^{\beta}
-
\frac{\Gamma\left(N_{\nu}-\frac{d}{2}-1\right)}
{\Gamma\left(N_{\nu}-\frac{d}{2}\right)}~F~\frac{g^{\alpha\beta}}{2} , \textrm{~~etc.,}
\eea
with $P_i^{\alpha}$ being the so-called chords introduced in (\ref{di1}).
For the general case $P_1(T)$ see section \ref{sec:onelnum}.

One now has to perform the $x$-integrations.
In  \ar{}{}, we will do this by the following simple formula: 
\bqa
\label{a1}
\int_0^1 \prod_{i=1}^N dx_i ~ x_i^{q_i-1}
~ \delta\left(1-\sum_j x_j\right)
&=&
\frac{\Gamma(q_1) \cdots \Gamma(q_N)}
{\Gamma\left(q_1 + \cdots + q_N \right)} .
\eqa
From the above text it is evident that the integrand of  (\ref{eq-scalar1}) contains besides simple sums of monomials $\prod_i x_i^{n_i}$ also different structures.
This is due to the appearance of the factors  $U(x)$ and $F(x)$.
Beginning with two-loop tensor integrals, one faces additionally a complicated dependence of $P(T)$ on $x$ for higher rank tensors $T$ due to the appearence of $V$ and $\alpha$, see (\ref{eq:plt2}).

For this reason, the present version of \ar{} is restricted to 
scalar and vector integrals and/or to one-loop integrals.
In these cases one may rewrite $F(x)$ and $U(x)$ so that (\ref{a1}) becomes applicable; for the one-loop case only the $F(x)$.
That is why we discuss here only the $F(x)$.
From (\ref{f1loop}), the $F(x)$ may be written as  a sum of $N_F \leq \frac{1}{2}N(N+1)$ non-vanishing, bilinear terms in $x_i$:
\bea\label{fxrep} 
F(x)^{-(N_\nu-dL/2)} &=& \left[ \sum_{n=1}^{N_F} f_n(i,j) x_i x_j \right]^{-(N_\nu-dL/2)}
\nl
&=& \frac{1}{\Gamma(N_\nu-dL/2)}\frac{1}{(2\pi i)^{N_F}}
\prod_{i=1}^{N_F} \int\limits_{-i \infty + u_i}^{i \infty + u_i} d z_i \prod_{n=2}^{N_F}
\left[ f_{n}(ij) x_i x_j\right] ^{z_n} 
\nl&&~
\left[ f_{1}(ij)x_i x_j\right] ^{-(N_\nu-dL/2)-\sum_{j=2}^{N_F} z_j} 
\Gamma\left( N_\nu-\frac{dL}{2}+\sum_{j=2}^{N_F}z_j\right) 
\prod_{j=2}^{N_F} \Gamma(-z_j) .
\nl
\eea
Here, $f_n(i,j) = f_{ij} $ if $f_{ij} \neq 0$. 
Inserting (\ref{fxrep}) (and if needed a similar representation for the $U(x)$) and the tensor function $P(T)$ into (\ref{eq-bha}) allows to apply (\ref{a1}) for an evaluation of the $x$-integrations. 

As a result,  any scalar Feynman integral may be represented by a single multi-dimensional MB-integral and $L$-loop tensor integrals by finite sums of MB-integrals.
With \ar{} 
we will evaluate  the $L$-loop integrals by a loop-by-loop technique, which essentially allows us to restrict the formalism to the one-loop case.
By the examples it will be seen that this is a powerful ansatz for many applications.   

In subsequent steps, the package \mb{} may be called.
This package needs as input some MB-integral(s), e.g. as being prepared by \ar.
As described in detail in \cite{Czakon:2005rk}, \mb{} allows to analytically expand a Feynman integral in $\eps$ and to evaluate the resulting sequence of finite MB-integrals by one or the other method.

\section{\label{sec:ambre}Using \ar}
In this section we describe the use of the 
Mathematica package \ar{}.  \ar{} stands for {\bfseries{A}}utomatic {\bfseries{M}}ellin-{\bfseries{B}}arnes {\bfseries{Re}}presentation. 
It is a (semi-)automatic procedure written for multi-loop calculations. 
The package works with Mathematica 5.0 and later versions of it.

The algorithm to build up MB-representations for Feynman integrals as described in the last section
consists of the following parts:
\begin{enumerate}
\item[(i)] define kinematical invariants which depend on the external momenta;
\item[(ii)] make a decision about the order in which $L$ 1-loop subloops 
$(L \geq 1)$  will be worked out sequentially;
\item[(iii)] construct a Feynman integral for the chosen subloop and 
perform manipulations
on the corresponding $F$-polynomial to make it optimal for later use of the MB 
representations;
\item[(iv)] use equation (\ref{fxrep});
\item[(v)] perform the integrations over Feynman parameters with equation (\ref{a1});
\item[(vi)] go back to  step (iii) and repeat the steps for the next subloop until $F$ in 
the last, $L^{th}$ subloop will be changed into an MB-integral.
\end{enumerate}
\begin{equation}
\label{alg}
\end{equation}
The steps (ii) and (iii) 
must be analyzed carefully, because there exists some freedom of choice
on the order of loop integrations in step (ii) and also
on the order of MB integrations in  
step (iii). 
Different choices may lead to different forms of MB-representations.

The present version 1.0 of \ar{} can be used to construct planar Mellin-Barnes representations for:
\begin{itemize}
  \item scalar multi-loop, multi-leg integrals
  \item tensor one-loop integrals 
  \item integrals with specific higher-rank numerators ending up with a single MB-integral
\end{itemize}
In the next sections several examples will be used for an introduction to specific features of the package.

Here,
we describe basic functions of the package. 
The starting point of all calculations is a proper definition of the integral (\ref{eq-bha}) and of the kinematical invariants to be used. 
Formally, it has to be done in the following way:
\begin{eqnarray}
  &&\verb+ Fullintegral[{numerator}, {denominator}, {internal momenta}]; +
\nonumber\\
  &&\verb+ invariants = {invariants as a rule};+
\label{FI}
\end{eqnarray}
We recommend to use $k_{i}$ and $p_{i}$ as symbols for internal and external momenta, respectively. 
Also non-zero masses should appear as symbols; a numeric value may cause problems in multi-loop calculations.

The command \verb+Fullintegral+ defines a given integral.
For example:
\begin{equation}
  \verb#  Fullintegral[{1}, {PR[k1, 0, n1]*PR[k1 + p1, m, n2]}, {k1}]; #
\end{equation}
corresponds to:
\begin{equation}
  \int \frac{d^{d}k_{1}}{i\pi^{d/2}}  \frac{1}{(k_{1}^2)^{n_{1}}[(k_{1}+p_{1})^2-m^2]^{n_{2}}}.
\end{equation}
The last argument in the \verb+Fullintegral+ function is a list of internal momenta. 
The order of internal momenta in this list controls the ordering of integrations (if iterated). 
For example \verb+{k3,k2,k1}+ defines the first integration to be over $k_{3}$, the second  over $k_{2}$ and the third over $k_{1}$. 
The next step is to prepare a subloop of the full integral by collecting all propagators which carry a given loop momentum $k_i$. 
We do this by initiating the consecutive functions:
\begin{equation}
  \verb+  IntPart[iteration] +
\end{equation}
Each iteration, $i=1,2,\dots,L$, prepares the appropriate subloop for the integration over the corresponding internal momentum. 
It will display a piece of the \verb+Fullintegral+ with:
\begin{itemize}
  \item the numerator associated with the given subloop
  \item subloop for a given internal momentum
  \item internal momentum for which \ar{} will integrate the subloop
\end{itemize}
The execution of \verb+IntPart[iteration]+ proceeds in the order  \verb+IntPart[1]+,  \verb+IntPart[2]+,  then \verb+IntPart[3]+, and so on. 
If there is a need to change the ordering of integrations, one has to change the order in the starting list of internal momenta (\ref{FI}).
Inserting \verb+IntPart[2]+ before \verb+IntPart[1]+ would not be a proper way to do this. 
In the output of \verb+IntPart[iteration]+ a tag message will be displayed:
\begin{eqnarray}
  &&\verb+ Fauto::mode: U and F polynomials will be calculated +
\nonumber\\
  &&\verb+ in AUTO mode. In order to use MANUAL mode execute Fauto[0]. +
\end{eqnarray}  
By running \verb+Fauto[0]+, \ar{} will calculate the $F$-polynomial (with name \verb+fupc+) for a given subloop. 
At this stage, a user might wish to modify \verb+fupc+ manually, e.g. by applying some changes in kinematics.

During the calculations, the \verb+FX+ function of \ar{} may appear in the $F$-polynomial. This function collects full squares of sums of Feynman parameters,  e.g.:
\bea
 \verb# FX[X[1]+X[3]]^2# ~~&\equiv&~~ (x_1+x_3)^2.
\eea
Such terms appear in the $F$-polynomials if some masses in the loops are equal.
They will later allow to apply Barnes' lemma leading to lower dimensional MB-representations.
At the other hand, the exponent two of the square may lead to arguments of $\Gamma$-functions in (\ref{fxrep}) with doubled integration variables, with  far-reaching consequences for an  analytical evaluation when a sum over an infinite series of residua is tried. 

The basic function for deriving the Mellin-Barnes representation is:
\begin{equation}
  \verb+ SubLoop[integral]+
\end{equation}
This function takes output generated by \verb+IntPart[iteration]+ and performs the following calculations:
\begin{itemize}
  \item calculate the $F$-polynomial for the subloop (only if \verb+Fauto[1]+ is set)
  \item determine the MB-representation for the $F$-polynomial
  \item integrate over Feynman parameters $x_i$  
\end{itemize}
As a result, the MB-representation for a given subloop integral will be displayed. 
In multi-loop calculations one will notice additional propagators (marked in red in the output of \ar) which appear from the intermediate $F$-polynomial (see section \ref{sec:2looppl} for an instructive example).

As mentioned, \ar{} can construct Mellin-Barnes representations for general one loop tensor integrals. 
The procedure of calculating such cases is basically the same, with few
minor differences. 
First of all, the numerator input must be defined.
A one-loop box diagram with numerators $(k_1p_1) (k_1p_2) (k_1p_3)$ 
might look like this:
\begin{eqnarray}
&&\verb#Fullintegral[#
\nonumber\\
&&\verb#   {k1*p1,k1*p2,k1*p3},#
\nonumber\\
&&\verb#   {PR[k1,m,n1]PR[k1+p1,0,n2]PR[k1+p1+p2,m,n3]PR[k1+p3,0,n4]},{k1}];#
\nl
\end{eqnarray}
We have written this procedure such that numerators consist of scalar products of internal and external momenta. 
In the calculations with tensors, the definitions of momentum flows in the subloops play a crucial role for the results and have to be controlled carefully.
Another difference  to scalar cases is the way how \ar{} displays results. 
Because they can be long, we decided to use a short notation. 
For example:
\begin{equation}
   \verb+{ARint[1],ARint[2],ARint[3]}+
\end{equation}
The result of the evaluation has to be understood as the sum of the elements,  
\\
\verb#ARint[1]+ARint[2]+ARint[3]#, 
\\
where each \verb+ARint[i]+ is one of the resulting MB-integrals. 
By executing 
\\
\verb+ARint[result,i]+ 
\\
one may display the appropriate \verb+ARint[i]+. 
The procedure uses the short notation by default, but it is also possible to use the option \verb+Result->True+ in order to force \verb+SubLoop+ to display the full result:
\begin{equation}
   \verb+SubLoop[integral,Result->True];+
\end{equation}
Finally, 
we have also implemented Barnes' first lemma:
  \begin{eqnarray}\label{barnes1} 
    \int_{-i \infty}^{i \infty} dz  \Gamma(a+z) \Gamma(b+z)
    \Gamma(c-z) \Gamma(d-z) &=&
    \frac{\Gamma(a+c)\Gamma(a+d)\Gamma(b+c)\Gamma(b+d)}{\Gamma(a+b+c+d)} ,
  \end{eqnarray}
and Barnes' second  lemma:
  \begin{eqnarray}\label{barnes2} 
    \int_{-i \infty}^{i \infty} &dz& \; \frac{\Gamma(a+z) \Gamma(b+z)
    \Gamma(c+z) \Gamma(d-z) \Gamma(e-z)}{\Gamma(a+b+c+d+e+z)} =
    \nonumber \\ &&
    \frac{\Gamma(a+d)\Gamma(a+e)\Gamma(b+d)\Gamma(b+e)\Gamma(c+d)\Gamma(c+e)}{
    \Gamma(a+b+d+e)\Gamma(a+c+d+e)\Gamma(b+c+d+e)}.
  \end{eqnarray}
The usage of Barnes' lemmas is simple; one has to execute:
\begin{equation}
   \verb+BarnesLemma[representation,i]+
\end{equation}
where \verb+i+ is $1$ or $2$ for the first or second Barnes' lemma, respectively. 
This function tries to apply the lemma on all  integration variables $z_{i}$ of the MB-representation which do not appear in the exponents  of kinematical invariants. 
It
also searches in  the exponents of kinematical invariants for pairs of integration variables. 
For example, be in one
 exponent the combination $(z_1 + z_2)$ and in another one the 
 combination $(-z_1-z_2)$.
This might appear as dictated by the structure of equation (\ref{mb}).
 The automatic change $z_1->z_1-z_2$  eliminates $z_2$ in these
 exponents
so that Barnes' lemma can be applied for $z_2$.
A  comment will be displayed if the lemma was successfully applied.
Barnes' first lemma is quite often applicable, while Barnes' second  lemma applies only sporadically (see {\ttfamily example14.nb}).
The automatic change of variables may be switched off by calling \verb+shift[0]+.

In the appendix we list the Mathematica functions of \ar.
\section{\label{sec:one-loop0}One-loop integrals}
We will give a couple of examples starting with construction of
MB-representations for the 1-loop Feynman integrals which are 
an important ingredient of the algorithm (\ref{alg}). 
Most of the cases considered in subsequent sections are connected 
with massless gauge theories or massive QED. 
\subsection{\label{sec:one-loop}Example: the pentagon diagram of massive QED}

\begin{figure}[tbhp]
  \begin{center}
    \epsfig{file=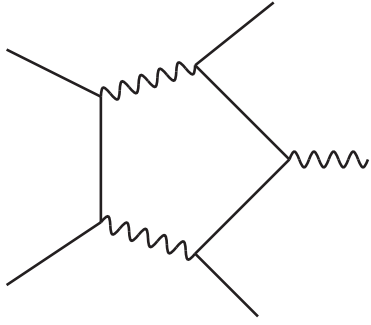,width=5cm}
  \end{center}
  \caption[Massive QED pentagon diagram (example1.nb, example2.nb)]{\label{5pf} Massive QED pentagon diagram}
\end{figure}

Let us consider  the one-loop five-point function shown 
in figure \ref{5pf}. 
The external momenta fulfill $p_3^2=0$, $p_i^2 = m^2$ for the other particles,  and the  $s_{ij} = (p_i+p_j)^2$ are kinematical invariants of the process.
If we naively use the 
\verb+FUPolynomial+ 
function of the \mb{} package, we will get:
\begin{eqnarray}
U&=& x_1 + x_2 + x_3 + x_4 + x_5, \label{u1}\\
F&=& 
m^2 x_1^2 + 2 m^2 x_1 x_3 - s_{15} x_1 x_3 + m^2 x_3^2 + 
 2 m^2 x_1 x_4 - s_{23} x_1 x_4 + m^2 x_2 x_4 \nonumber \\
&-& s_{45} x_2 x_4 + 
 2 m^2 x_3 x_4 + m^2 x_4^2 - s_{12} x_2 x_5 + m^2 x_3 x_5 - 
 s_{34} x_3 x_5.
\label{f1}
\end{eqnarray}
A simple counting of terms in the $F$-polynomial 
would prove that this leads to a twelve-dimensional MB-integral. 
Of course the terms in $F$ can be grouped 
from the beginning and we will see in a minute 
that a five-fold MB-integral may be obtained; see also the sample file {\ttfamily example1.nb}\footnote{The sample Mathematica files are part of the package \ar. They are available at  \cite{Katowice-CAS:2007,Zeuthen-CAS:2004}.}. 

First, propagators and kinematical invariants are defined:
\begin{eqnarray}
&&\verb#Fullintegral[#
\nonumber\\
&&\verb#    {1},#
\nonumber\\ 
&&\verb#    {PR[k1 + p1, 1, n1]*PR[k1 + p1 + p5, 0, n2]*#
\nonumber\\
&&\verb#    PR[k1 + p1 + p4 + p5, 1, n3]*PR[k1 + p1 + p3 + p4 + p5, 1, n4]*#
\nonumber\\
&&\verb#    PR[k1 + p1 + p2 + p3 + p4 + p5, 0, n5]},{k1}];#
\end{eqnarray}
The kinematics is defined in a cyclic way:
\begin{equation}
p_i^2 = m_i^2,\qquad s_{i,i+1} = (p_i+p_{i+1})^2, \qquad i=1,...,5.
\label{5kin}
\end{equation}
Then, using the {\tt IntPart} and {\tt SubLoop} functions the steps 
(ii)-(v) of the algorithm (\ref{alg}) are worked out and 
we end up with a nine-fold MB-representation.
This representation is due to the following $F$-polynomial, constructed in 
the automatic way by \ar:
\begin{eqnarray}
F^{'}&=& m^2 (x_1 + x_3 + x_4)^2 - s_{15}x_1x_3 - s_{23}x_1x_4 + 
 m^2x_2x_4 - s_{45}x_2x_4 - s_{12}x_2x_5 
\nonumber \\
&& +~m^2x_3x_5 - 
 s_{34}x_3x_5.
\label{fprim}
\end{eqnarray}
Some mass terms have been collected here, but
the $F$-polynomial can be further simplified 
by redefining 
$s_{34} \to \bar{s}_{34}+m^2$ and $s_{45} \to \bar{s}_{45}+m^2$,
so that each term $x_ix_j$ appears only once.
The $F^{'}$ polynomial becomes finally: 
\begin{eqnarray}
F^{''}&=& m^2 (x_1 + x_3 + x_4)^2 - s_{15}x_1x_3 - s_{23}x_1x_4  
- \bar{s}_{45}x_2x_4 - s_{12}x_2x_5 -  \bar{s}_{34}x_3x_5,
\label{fbis}
\end{eqnarray}
which gives a seven-fold MB-representation.
In certain cases, some of the MB-integrations do not depend on the kinematics and Barnes lemmas may be applied.
Here, due to the term $(x_1 + x_3 + x_4)^2$ one may twice use Barnes' first lemma (\ref{barnes1}) and thus the MB-representation can be further reduced to a five-fold integral.
A five-particle scattering process depends on five variables (plus a mass in Bhabha scattering), so a further simplification is impossible.

In sample file {\ttfamily example2.nb}, we use another definition of kinematical variables, 
namely 
\begin{equation}
\begin{array}{lll}
p_i^2 = m^2, & p_1    p_2 = \frac{1}{2}(t'-2 m^2 ), &
 p_1    p_3 = \frac{1}{2}(t - t' - v_1), \\
 p_1    p_4 = m^2 + \frac{1}{2}(v_1-s - t), &
 p_1    p_5 = \frac{1}{2}(s-2 m^2), &
p_2    p_3 = \frac12 v_1,   \\
 p_2    p_4 = \frac12(s - v_1 - v_2-2 m^2), &
p_2    p_5 = \frac12(v_2-s - t'+2 m^2), & \\ 
 p_3    p_4 = \frac12 v_2, & 
p_3    p_5 = \frac12 (t'-t - v_2), & 
p_4    p_5 = \frac12 (t-2 m^2),
\end{array}
\label{invA}
\end{equation}
and we get  $F$ directly in the form:
\begin{eqnarray}
F^{'''}&=&(x_1 + x_3 + x_4)^2 - t x_1 x_3 - t' x_1 x_4 - v_2 x_2 x_4 - 
  s x_2 x_5 - v_1 x_3 x_5.
\end{eqnarray}
No wonder, that using function {\tt SubLoop} we obtain directly the smallest,
seven-dimensional integral, which then again reduces to the five-fold 
integral. 
The resulting MB-represen\-ta\-tion for the scalar Feynman integral is:
\bea\label{mbi}
G[1] &=&
\frac{-e^{\eps\gamma_E}}{(2\pi i)^5}
\prod_{i=1}^5 \int_{-i\infty+u_i}^{+i\infty+u_i} d r_i
 (-s)^{-3-\eps-r_1} 
(-t)^{r_2} (-t')^{r_3} \left(\frac{v_1}{s}\right)^{r_4} \left(\frac{v_2}{s}\right)^{r_5}
\nl&&
\Gamma[-r_2] \Gamma[-r_3] \Gamma[1 + r_2 + r_3] \Gamma[-r_1 + r_2 + r_3]
\Gamma[-2-\eps - r_1 - r_4]\Gamma[-r_4]
\nl&&
\Gamma[1  + r_2 + r_4]\Gamma[ -2-\eps - r_1 - r_5]
\Gamma[-r_5]\Gamma[1  + r_3 + r_5]
\nl&&
\frac{
\Gamma[3+\eps +  r_1 + r_4 + r_5]
\Gamma[3  + 2 r_1 + r_4 + r_5]
}
{
\Gamma[-1-2\eps] \Gamma[3  + 2 (r_2 + r_3) + r_4 + r_5]
}.
\eea
The real parts  of the integration strips are $-2 < u_1 < -1$ and $-1/2 < u_i < 0 , i=2 \ldots 5$.

A subsequent  application of \mb{} shows that
up to constant terms in $\epsilon$, needed for an evaluation of two-loop
massive Bhabha scattering \cite{Fleischer:2007xx}, there are maximally three-dimensional finite contributions to be evaluated further. 

\vfill

\subsection{\label{sec:onelnum}Numerators}
\ar{} may handle arbitrary one-loop tensor integrals.
The one-loop Feynman parameter integral for a tensor of degree $m$ is the generalization of equation (\ref{eq:p1tens2}):
\begin{eqnarray}\label{eq:tensorAnast}
G_1(T_m)&\equiv&
G_1(k^{\mu_1}\cdots k^{\mu_m})
\nl
&=&
\frac{(-1)^{N_{\nu}}}{\prod_{i=1}^N \Gamma(\nu_i)} 
\int
\prod_{i=1}^{N}dx_i x_{i}^{\nu_{i}-1} \delta(1-\sum_{j=1}^N x_j)
\sum_{r=0}^m 
\frac{\Gamma\left(n-\frac{d+r}{2}\right)}{ (-2)^{\frac{r}{2}} F^{n-\frac{d+r}{2}} }
\left\{{\cal A}_r P^{m-r}\right\}^{[\mu_1,\ldots,\mu_m]} .
\nl
\end{eqnarray}
Here $F\equiv F(x)$ and $P \equiv P_1(k^{\mu}) = \sum_i x_i P_i = \sum_{i,e}x_i\beta_{ie}p_e^{\mu}$ were introduced in equations (\ref{eq:F}) and (\ref{eq:p1k}).
The $r$ starts from zero (with ${\cal A}_0=1$), and it is ${\cal A}_r=0$ for $r$ odd, and 
${\cal A}_r= g^{\left[ \mu_{i_1} \mu_{i_2}\right.}\cdots g^{\left. \mu_{i_{r-1}} \mu_{i_r}\right]}$
 for $r$ even. 
The convention $[\mu_{i_1} \ldots]$ means the totally symmetric combination of the arguments.
\\
In \ar{}  tensorial numerators are assumed to be contracted with the external momenta $p_e$, so that the following quantity is evaluated:
\bea
P_m ~ G_1(T_m) &\equiv& \left( p_{e_1}^{\mu_1}\cdots p_{e_m}^{\mu_m} \right)~ G_1(k^{\mu_1}\cdots k^{\mu_m}).
\eea

\begin{figure}
  \begin{center}
    \epsfig{file=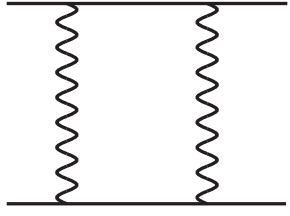,width=4cm}
  \end{center}
  \caption[Massive QED one-loop box diagram (example3.nb)]{\label{fig:1loopbox} Massive QED one-loop box diagram}
\end{figure}

As an example, we have prepared the massive QED one-loop box of figure \ref{fig:1loopbox} in sample file {\ttfamily  example3.nb}
with the numerator
$	(k_{1}p_{1})(k_{1} p_{2})(k_{1} p_{3})$.
The corresponding definition used in \ar{} 
is:
\begin{eqnarray}
&&\verb#Fullintegral[#
\nonumber\\
&&\verb#   {k1*p1,k1*p2,k1*p3},#
\nonumber\\
&&\verb#   {PR[k1,m,n1]PR[k1+p1,0,n2]PR[k1+p1+p2,m,n3]PR[k1+p3,0,n4]},{k1}];#
\nl
\end{eqnarray}
Obviously, when working with tensor integrals we expect the result to be 
a sum of several MB-integrals (the higher the rank is, the more integrals will 
be obtained). 
We have cross checked numerically results for two-, three- and four-point 
functions by comparing our results (from using \ar{} and \mb) with decompositions of integrals into master 
integrals using the {\bfseries{I}}ntegration-{\bfseries{B}}y-{\bfseries{P}}arts method implemented in the package {\ttfamily IdSolver} (M. Czakon, unpublished).  
Cross checks were done for numerators with up to eight scalar products in the numerators of the Feynman integrals. 

Finally we refer to section \ref{subsec:multinum} for the interesting special case of irreducible numerators arising in intermediate subloops. 
In certain cases, the result for a tensor integral  may remain as compact as it is for scalar integrals.  
\subsection{\label{subs:masses}More masses}
\begin{figure}
  \begin{center}
    \epsfig{file=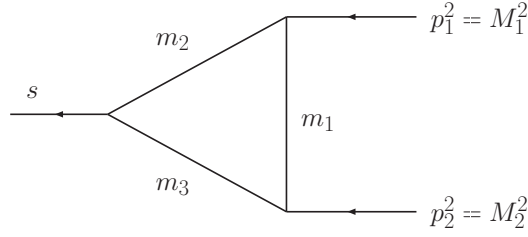,width=7cm}
  \end{center}
  \caption[General one-loop vertex (example4.nb)]{\label{vg}
General one-loop vertex}
\end{figure}

$N$-point functions with arbitrary internal masses and off-shell external
legs give complicated multi-dimensional MB-integrals.
Let us consider here and in {\ttfamily example4.nb} a general one-loop scalar vertex, Fig.~\ref{vg}. 
In this case we get a five-dimensional MB-integral:

\begin{eqnarray}
V_{\rm general} &=&
 \frac{(-1)^{n_{123}}}
{ \Gamma[n_1]  \Gamma[n_2]  \Gamma[n_3]  \Gamma[4 - 2 \eps  - n_{123}]}
\nonumber \\
&&  \frac{1}{(2\pi i)^5}
 \int_{-i \infty}^{+i \infty} d z_1
 \int_{-i \infty}^{+i \infty} d z_2
 \int_{-i \infty}^{+i \infty} d z_3
 \int_{-i \infty}^{+i \infty} d z_4
 \int_{-i \infty}^{+i \infty} d z_5
 \prod_{i=1}^5 \frac{\Gamma[-z_{i}]}{\Gamma[2-\epsilon-n_1-z_1+z_5]}
 \nonumber \\
&& {(m_1^2)}^{z_1} {(m_2^2)}^{z_3}
{(m_3^2)}^{2-\epsilon - n_{123} - z_{12345}} {(m_1^2+m_2^2-M_1^2)}^{z_2}
 {(m_1^2+m_3^2-M_2^2)}^{z_4} \nonumber \\
&& {(m_2^2+m_3^2-M_1^2-M_2^2-s)}^{z_5}  \Gamma[n_1 + z_{1124}]
   \Gamma[4 - 2 \eps - n_{1123} - z_{1124}]
\nonumber \\
&& \Gamma[n_2 + z_{235}]
   \Gamma[-2 + \eps  + n_{123} + z_{12345}],
\end{eqnarray}
where we abbreviated $z_{1124} = 2z_{1}+ z_{2}+ z_{4}$ and $n_{123} =
n_1+n_2+n_3$, etc.

For the massive QED case, $M_1=M_2=m_2=m_3=m, m_1=0$,  we get a compact
one-dimensional MB-representation: 
\begin{eqnarray}
V_{\rm QED} &=&
 \frac{(-1)^{n_{123}} \Gamma[4-2 \epsilon-  n_{1123}]}
{  \Gamma[n_2]  \Gamma[n_3] \Gamma[4-2 \epsilon-n_{123}]}  \frac{1}{2\pi i}
 \int_{-i \infty}^{+i \infty} d z {(m^2)}^{z}
{(-s)}^{2-\epsilon-n_{123}-z}
   \nonumber \\
&& \frac{\Gamma[-z] \Gamma[2-\epsilon-n_{12} - z] \Gamma[2-\epsilon-n_{13}
- z]
   \Gamma[-2 + \epsilon + n_{123} + z] }
{ \Gamma[4 - 2 \eps - n_{1123}-2 z]}.
\end{eqnarray}

\subsection{\label{subs:legs}More legs}
For topologies with a higher number of legs, there is an increasing number of
kinematical invariants and so the dimension of MB-representations increases. 
The number of dimensions may become smaller after analytical continuation in $\eps$ and for lowest orders in $\eps$. 
For a scalar or vector Bhabha massive five-point function, 
Fig.~\ref{5pf}, up to constant terms in $\epsilon$, it includes
at most three-dimensional integrals, 
which hopefully can be solved even analytically \cite{Fleischer:2007xx}. 
In general, the MB-representation for that case is five-dimensional, see section \ref{5pf}.

\begin{figure}[t]
  \begin{center}
    \epsfig{file=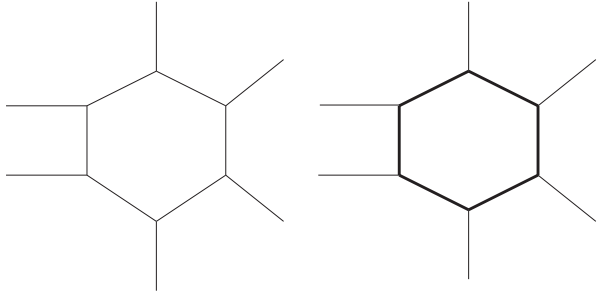,width=8cm}
  \end{center}
  \caption[Six-point scalar functions  (example5.nb); left: massless case,
right:  massive case]{\label{6pf}
Six-point scalar functions; left: massless case,
right:  massive case}
\end{figure}

In {\ttfamily example5.nb} we derive  MB-representations for a massless and a massive one-loop hexagon scalar  diagram, see figure \ref{6pf}. 
In general, it is an eight-fold integral, but the constant term in $\epsilon$
includes again only up to three-dimensional MB-integrals.

If all internal lines have equal non-vanishing mass, one has to deal with 
a nine-dimensional MB-integral.
Again, the numerical results have been checked for both cases in the 
Euclidean region against sector decomposition. 
The package contains the auxiliary file 
\\
{\ttfamily KinematicsGen.m} which generates the kinematics for six-point functions with arbitrary off-shell external legs.

\section{\label{sec:beyond}Multi-loop integrals: loop-by-loop integrations }
The Feynman integral (\ref{eq-scalar1}) includes
a delta-function which makes $U = 1$ for one-loop diagrams there so that the MB-relation (\ref{mb}) acts only on $F$.
 This simplification can be made also useful in multi-loop integrals by performing loop-by-loop integrations.
We collected few examples which will exhibit several specific features.

\subsection{\label{sec:2looppl}Example: two-loop planar box in massive QED}
Let us take the massive two-loop planar box topology\footnote{In fact there are three double-box diagrams in massive QED. One of them is non-planar, and we discuss here the so-called first planar diagram \cite{Smirnov:2001cm}.} with seven internal lines  as introduced in {\ttfamily example6.nb}.
The momentum flow is defined in the following way, with all momenta being incoming:

\begin{eqnarray}
  &&\verb#Fullintegral[#
\nonumber\\
  &&\verb#    {1},#
\nonumber\\
  &&\verb#    {PR[k1, m, n1]PR[k1 + p1, 0, n2]PR[k1 + p1 + p2, m, n3]#
\nonumber\\
  &&\verb#    PR[k1 - k2, 0, n4]PR[k2, m, n5]PR[k2 + p1 + p2, m, n6]#
\nonumber\\
  &&\verb#    PR[k2 - p3, 0, n7]}, {k2, k1}]#.
\label{2loopbasic}
\end{eqnarray}
First, the momentum integration over $k_2$ is taken.
The  $k_2$ flow in the first 
subloop is defined by the function \verb+IntPart[1]+, which contains all propagators with  momentum $k_2$: 
\begin{eqnarray}
  &&\verb#integral = PR[k1 - k2, 0, n4]*PR[k2, m, n5]*#
\nonumber\\
  &&\verb#           PR[k2 + p1 + p2, m,n6]*PR[k2 - p3, 0, n7]#.
\end{eqnarray}

We just mention that generally it is preferred to choose the order of iteration such  
that first  the loops with lowest number of lines 
are executed.
Then their $F$-polynomials have a minimal number of terms. 
The first loop's $F$-polynomial is the \verb+SubLoop[integral]+ function:

\begin{figure}[t]
  \begin{center}
    \epsfig{file=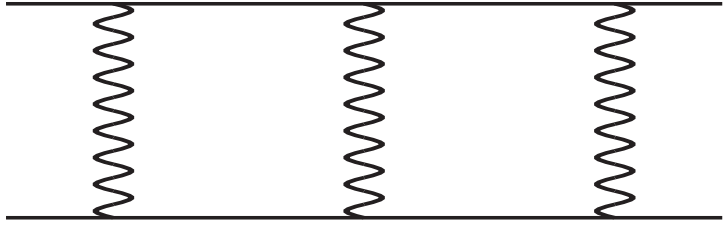,width=10cm,height=3cm}
  \end{center}
  \caption[Massive two-loop planar QED box (section \ref{sec:2looppl}: example6.nb, section \ref{subsec:multinum}: example7.nb)]
{\label{b1} Massive two-loop planar QED box}
\end{figure}

\begin{eqnarray}
&&F[k_2]\equiv \verb#fupc=#
\nonumber\\  
&&\verb#         m^2*FX[X[2] + X[3]]^2 - PR[k1, m]*X[1]*X[2]-#
\nonumber\\
&&\verb#         PR[k1 + p1 + p2, m]*X[1]*X[3] - s*X[2]*X[3] -#  
\nonumber\\
&&\verb#         PR[k1 - p3, 0]*X[1]*X[4] + 4*m^2*X[3]*X[4] -# 
\nonumber\\
&&\verb#         s*X[3]*X[4] - t*X[3]*X[4] - u*X[3]*X[4]#
\label{b1f}
\end{eqnarray}
It is reproduced here as derived without interactions by the user.
The $F$-polnomial contains a mass term with  the {\ttfamily FX} function which later will allow to apply Barnes' first lemma successfully, and also a redundancy in  {\ttfamily X[3]*X[4]}.
The following nine-fold MB-representation after integrating over $k_2$ is obtained:
\begin{eqnarray}
  &&\verb#SubLoop1[((-1))^(n4 + n5 + n6 + n7 + z2 + z3 + z5) 4^z6#
\nonumber\\
  &&\verb#(m^2^(z1 + z6) (-s)^(z4 + z7) (-t)^z8#
\nonumber\\
  &&\verb#(-u)^(2-ep - n4-n5-n6-n7 - z1 - z2 - z3 - z4 - z5 - z6 - z7 - z8)#
\nonumber\\
  &&\verb#Gamma[-z1] Gamma[(-z2)] Gamma[(-z3)] Gamma[#
\nonumber\\
  &&\verb#2 - ep - n4 - n5 - n6 - z1 - z2 - z3 - z4] Gamma[(-z4)]# 
\nonumber\\
  &&\verb#Gamma[(-z5)] Gamma[n4 + z2 + z3 + z5] Gamma[(-z6)] Gamma[(-z7)]# 
\nonumber\\
  &&\verb#Gamma[-z8] Gamma[-2+ep + n4 + n5 + n6 + n7 + z1 + z2 + z3 + z4 + z5 +# 
\nonumber\\
  &&\verb#z6 + z7 + z8] Gamma[ 2 - ep - n4 - n5 - n7 + z1 - z2 - z5 - z9]#
\nonumber\\
  &&\verb#Gamma[(-z9)] Gamma[(-2) z1 + z9] Gamma[n5 + z2 + z4 + z9])/#
\nonumber\\
  &&\verb#(Gamma[n4] Gamma[n5] Gamma[n6] Gamma[4 - 2 ep - n4 - n5 - n6 - n7]#
\nonumber\\ 
  &&\verb#Gamma[n7] Gamma[(-2) z1])),# 
\nonumber\\
  &&\verb#PR[k1, m, z2]PR[k1 + p1 + p2, m, z3]PR[k1 - p3, 0, z5]])#
\label{it1}
\end{eqnarray}
It is clear that 
the factors in front of the \verb+X[3]X[4]+ coefficient sum up to zero, due to $s+t+u = 4 m^2$.
To remove them from the beginning, the \verb+Fauto[0]+ option must be executed,
followed by a modification of $F$: 
\begin{equation}
\verb# fupc = fupc /. u -> 4*m^2-s-t# .
\label{repl}
\end{equation}
In this way, executing the \verb+SubLoop[integral]+ function again, the MB-representation 
becomes five-dimensional, and also the term $4^{z_6}$ is absent now. 

The same situation appears in the second iteration, when integrating over $k_1$. 
We can switch to the \verb+Fauto[0]+ mode and again modify $F$.
After again applying Barnes' first lemma, 
we end up with a six-dimensional integral.

Of course, by writing from the very beginning the kinematical invariants 
without the invariant $u$, one can 
work out the whole case fully automatic with mode \verb+Fauto[1]+.
\subsection{\label{subsec:multinum}Special numerators }
The {\ttfamily example6.nb} is interesting in yet another respect.
After the first integration, the propagators for the second one
contain four propagators, some of them with shifted indices compared to the input:
 \bea
\verb#    PR[k1, m, n1-z2]PR[k1 + p1, 0, n2]PR[k1 + p1 + p2, m, n3-z3]#
\nonumber\\
  \verb#PR[k1 - p3, 0, z5]# .
\label{it1b}
\eea
This corresponds to the one-loop box of example {\ttfamily example3.nb} discussed in section \ref{sec:onelnum}, but with   shifted indices.
It includes the one new propagator with momentum $q_5 = k_1 - p_3$.
If we would have been evaluating an integral with numerator $(q_5^2)^{-n_8}$ and 
repeat the calculation, we would get after the $k_2$ integral
an $F$-polynomial with one of 
the terms including the propagator \verb#PR[k1 + p1 + p2 + p4, 0, 1]#; see
\verb#SubLoop[integral]# in {\ttfamily example7.nb}; see also  \cite{Smirnov:2004}. 
It will sum up with \verb#PR[k1 + p1 + p2 + p4, 0, -n8]#
resulting in the following integral
\begin{eqnarray}
&&\verb#integral= PR[k1, m, n1 - z2] PR[k1 + p1, 0, n2]# 
\nonumber\\
&&\verb#          PR[k1 + p1 + p2, m, n3 - z3]# 
\nonumber\\
&&\verb#          PR[k1 + p1 + p2 + p4, 0, -2 + ep + n4 +#
\nonumber\\
&&\verb#             n5 + n6 + n7 - n8 + z1 + z2 + z3 + z4]# ,
\end{eqnarray}
which has the following well-known  $F$-form of the one-loop box: 
$$
\verb#m^2 FX[X[1] + X[3]]^2 - s X[1] X[3] - t X[2] X[4]#  .
$$
What is essential here, no additional momentum structure appears.

Analyzing the irreducible numerators of the topology for the given momentum choice, one finds that there are two scalar products which may not be represented by linear combinations of the propagators (and thus are called irreducible):  $k_1p_3$ together with $k_2p_1$ or $k_1p_3$ together with $k_2p_2$.
So, $q_5^2$ represents one of two existing irreducible numerators and it may be quite useful to have a simple MB-representation for that case.
We see that there are integrals with (selected) numerators which may be represented by a single MB-representation as if a scalar integral would have been studied.
This was used several times in examples given in \cite{Smirnov:2004,Smirnov:2006ry} and in references cited therein, and it was also used e.g. in \cite{Czakon:2006pa} for a study of massive two-loop box master integrals, and for more sophisticated four-loop cases in \cite{Bern:2006ew}.

Finally, a six-dimensional MB-integral emerges like in the scalar case.
To check this integral 
numerically with the \mb{}  package, two analytical continuations, one in $\epsilon$
and one in one of the powers of propagators must be done. 
We have checked
the numerical result also against the results we got from a
sector decomposition calculation and from a small-mass  expanded version \cite{Czakon:2004tg,Czakon:2006pa}.
\subsection{\label{subs:34l}Further examples: A three-loop planar box, a four-loop self-energy, and a two-loop pentagon}
\begin{figure}
  \begin{center}
    \epsfig{file=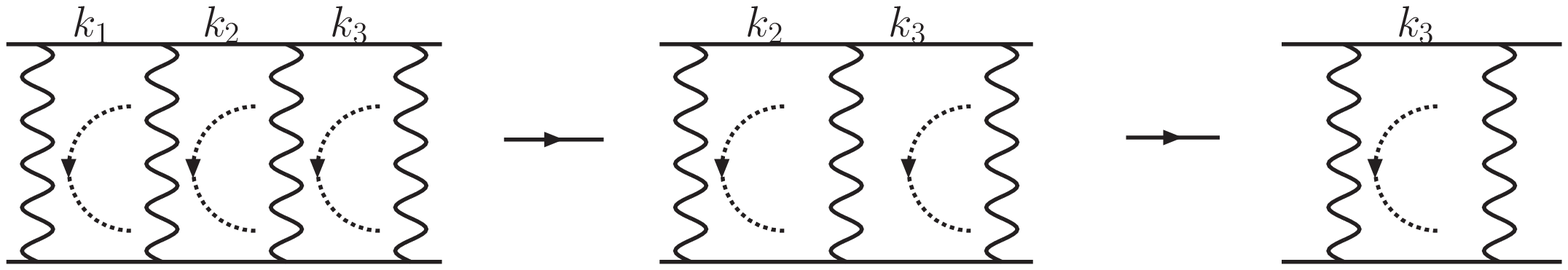,width=15cm,height=3cm}
  \end{center}
  \caption[The loop-by-loop iterative procedure 
(example8.nb)]{\label{P3l}
The loop-by-loop iterative procedure
}
\end{figure}

A three-loop planar integral, shown in figure \ref{P3l}, is treated in {\ttfamily example8.nb}.
The result is a 10-fold MB-representation.
With the \mb{} package it was shown 
that the numerical result agrees with \cite{Smirnov:2006ry}. 

The dimensions of some MB-representations for several massless and massive 
ladder topologies are summarized in Table~\ref{table}.
We apply an iterative procedure. 
For planar topologies the loop-by-loop iteration gives
always proper topologies which obey momentum conservation. 
Only some powers 
of propagators change into non-integer (complex) numbers. 

\begin{table}[b]
\begin{tabular}{rrrr|rrrr}
\hline \hline
Massless &&&& Massive &&& \\
\hline \hline
1-loop & 2-loop & 3-loop & 4-loop & 1-loop & 2-loop & 3-loop & 4-loop \\
  1    &   4    &   7    &   10   &   3    &   8    &   13   &   18   \\
  1    &   4    &   7    &   10   &   2    &   6    &   10   &   14   \\
\hline \hline
\end{tabular}
\caption{Dimensions of ladder topologies before and after applying Barnes' first
Lemma. 
}
\label{table}
\end{table}

A similar procedure can be applied to more complicated topologies which obey 
the same rule: integrating over an internal momentum leads to a topology with
propagators and momentum flow obeying momentum conservation in the remaining 
parts, i.e. we get regular subtopologies.

In this procedure, the choice of momenta flowing and the order of 
iterations are very important.
Look e.g. at the two-loop ladder example, also shown in figure \ref{P3l}.
If we would allow for the momentum flow $k_1$ through all  the outer lines, and take first
the integration over $k_1$ and then that over $k_2$, 
the final representation would not come out optimal (and Barnes' lemmas do not help).
Starting instead with the $k_2$ integration, we will again end up, as with the momentum flows shown in the figure, with a six-dimensional representation.

\begin{figure}
  \begin{center}
      \epsfig{file=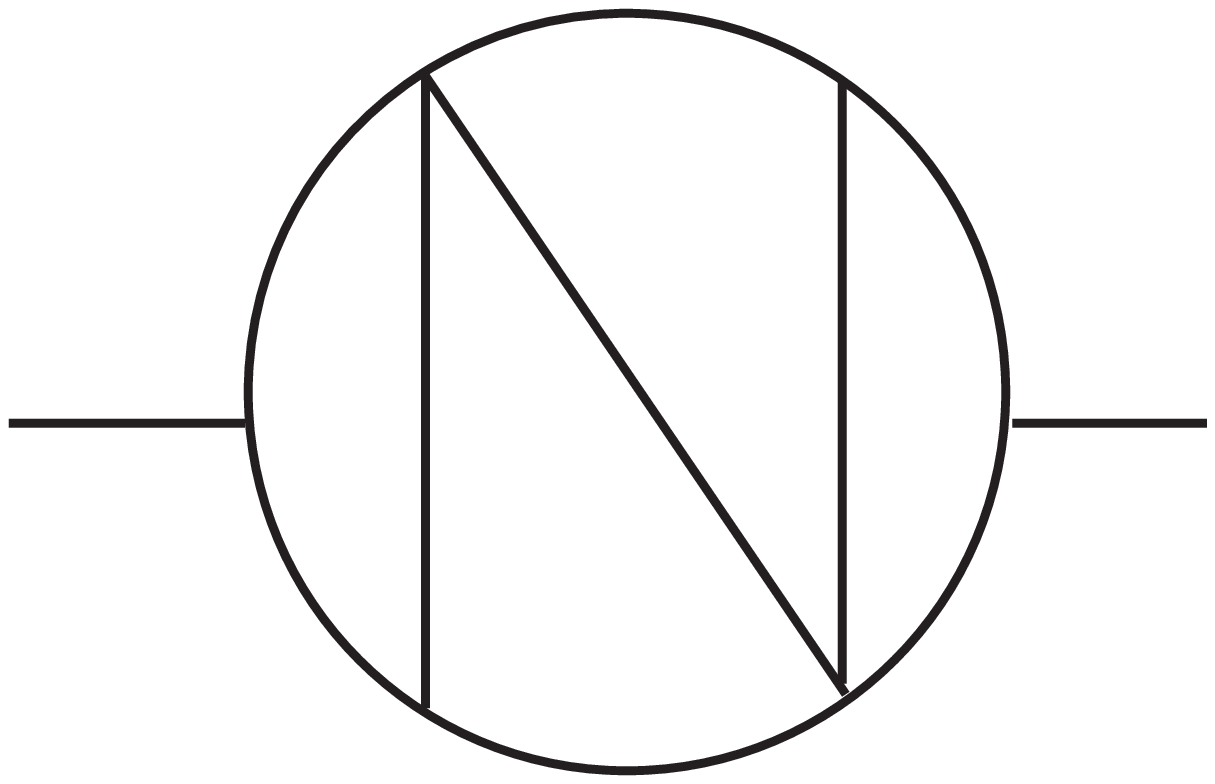,width=4.5cm,height=3.5cm}
\hspace*{2cm}      \epsfig{file=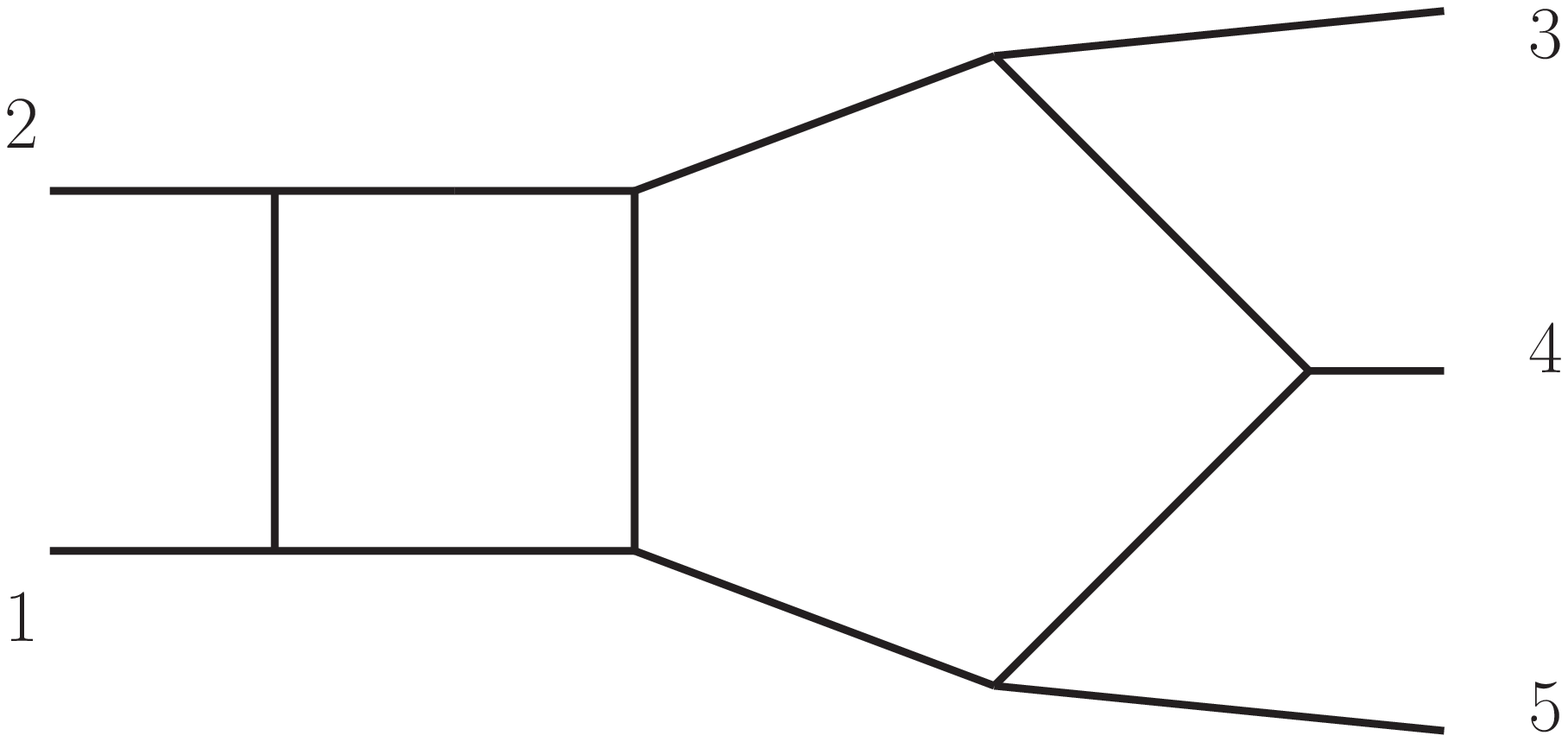,width=5.5cm,height=3.5cm}
  \end{center}
  \caption[Massless topologies; left: four-loop two-point diagram (example9.nb), right: two-loop five-point diagram (example10.nb)]{\label{P3L4}
Massless topologies; left: four-loop two-point diagram, right: two-loop five-point diagram}
\end{figure}

In files {\tt example9.nb} and {\tt example10.nb}, 
massless MB-representations are constructed for a four-loop two-point topology  
and for a two-loop five-point massless topology, see figure \ref{P3L4}.
The six-dimensional four-loop self-energy has been checked numerically against sector 
decomposition.
In {\ttfamily example10.nb}, there are three different derivations of MB-representations
for the same kinematics, defined by equation (\ref{5kin}). 
In each case we got another dimension of MB-integrals.
The minimal dimension of the integral is seven when we integrate first
over internal momenta of the box and then over that of the pentagon.
We checked that
 this agrees  numerically with \cite{Bern:2006vw} where also a
seven-dimensional MB-integral has been obtained.
If we integrate first over
the internal momentum running in the pentagon and next over that in the
box, then
a nine-dimensional MB-integral is obtained; again numerically they agree. 
In the third 
derivation, the momentum flow in the propagators is chosen
in a different way. 
Then a 13-dimensional MB-integral results.

\section{\label{sec:tadp}Tadpoles}
\begin{figure}
  \begin{center}
    \epsfig{file=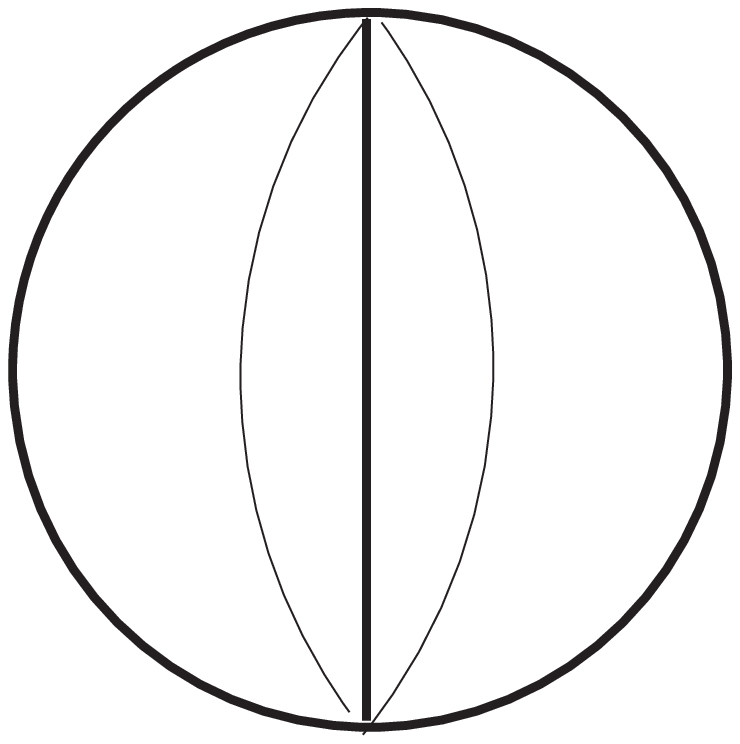,width=4cm,height=4cm}
\hspace*{1cm} 
    \epsfig{file=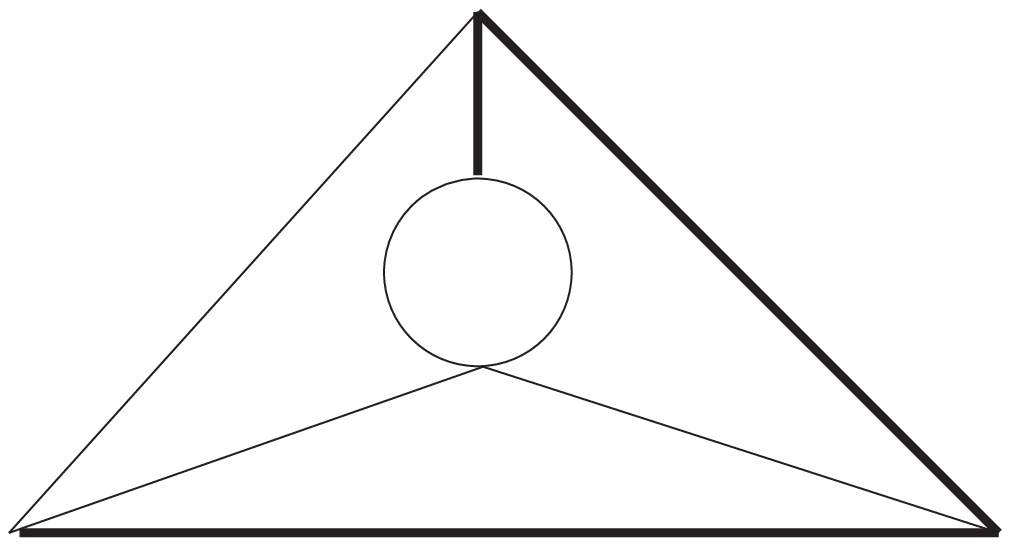,width=6cm,height=3.2cm}
   \end{center}
  \caption[Four-loop tadpoles with three massive lines; left: with a one-dimensional MB-representation (example11.nb), right: with a six-dimensional MB-representation (example12.nb)]{\label{ftadp}
Four-loop tadpoles with three massive lines; left: with a one-dimensional MB-representation, right: with a six-dimensional MB-representation
}
\end{figure}

The loop-by-loop approach  can also be applied to planar tadpoles.
Attention must be paid to keep the right order of integrations. 
Making iterations with the  {\ttfamily Fauto[1]} option (i.e. automatic), 
we may end up with three different  forms of propagators 
in the last iteration:
one massive propagator, massive and massless propagators, 
or one massless propagator. 
For the first situation the well known formula is used in \ar:
\begin{equation}
\int \frac{d^dk}{(k^2-q^2)^\nu} = i \pi^{d/2} (-1)^\nu \frac{
\Gamma [\nu+\epsilon-2]}{\Gamma[\nu]} \frac{1}{(q^2)^{\nu+\epsilon-2}}.
\end{equation}
We found that for some massive tadpoles  a term  $(-m)^\alpha$ can appear
which would lead to an oscillatory error while doing numerical calculations 
with \mb. 
In such a situation one has to go back to the previous 
subloop and modify the $F$-polynomial with {\ttfamily Fauto[0]} so that two propagators 
with equal momenta appear: a massive and a massless one. 
The same procedure must be applied when a single massless propagator 
appears in the last integral.
\\ 
We give as an example {\tt example11.nb}, for the diagram also shown left in figure \ref{ftadp}. 
Using \ar{},  we have constructed a one-dimensional Mellin-Barnes 
representation: 
\begin{eqnarray}
  T(n_1,...,n_5) &=&
  (-1)^{2+n_{12345}} (m^2)^{8-4 \epsilon-n_{12345}}
  \frac{1}{\prod_{i=1}^{5} \Gamma[n_i]}
  \frac{\Gamma[2- \epsilon-n_4] \Gamma[2-\epsilon-n_5]}
       {\Gamma[2- \epsilon]}
  \int_{-i \infty}^{+i \infty} d z_1
\nonumber\\
&\times& \Gamma[-z_1]
         \Gamma[2-\epsilon-n_1-z_1]
         \Gamma[2-\epsilon-n_2-z_1]
         \Gamma[4-2\epsilon-n_{12}-z_1]
\nonumber\\
&\times& \frac{
         \Gamma[-6+3\epsilon+n_{1245}+z_1]
         \Gamma[-8+4\epsilon+n_{12345}+z_1]}
         {\Gamma[4-2\epsilon-n_{12}-2z_1]}
\label{tadp}
\end{eqnarray}
At this point the proper order of integrations is very important. 
A different choice can lead to two- or even higher-dimensional representations.

Using \mb{} we got for the basic integral numerically: 
\begin{eqnarray}
T(1,1,1,1,1)&=&
  \frac{0.25}{\epsilon^4}  + \frac{1}{\epsilon^3} 
+ 2.843300366757447 \frac{1}{\epsilon^2} + 
 5.781543610421033 \frac{1}{\epsilon} \nonumber \\
&+& 22.955621881705923  
+ 80.89550616785341  \epsilon + 
 1085.2836587072804  \epsilon^2 \nonumber \\
&+&  4545.303884134432  \epsilon^3 + 
 35998.99383263255  \epsilon^4,
\end{eqnarray}
This is 
in agreement with \cite{Boughezal:2006xk}.

However, it appears that MB-representations for four-loop tadpoles
can be more complicated. 
In {\ttfamily example12.nb}, treating the diagram in figure \ref{ftadp} (right), we get a six-dimensional MB-integral. 
Taking into account other approaches \cite{Boughezal:2006xk,Faisst:2006sr}, one may
see that the MB-approach to multi-loop calculations has natural limits, 
especially in the massive cases.

\section{\label{sec:os}On-shell diagrams}
\begin{figure}
  \begin{center}
    \epsfig{file=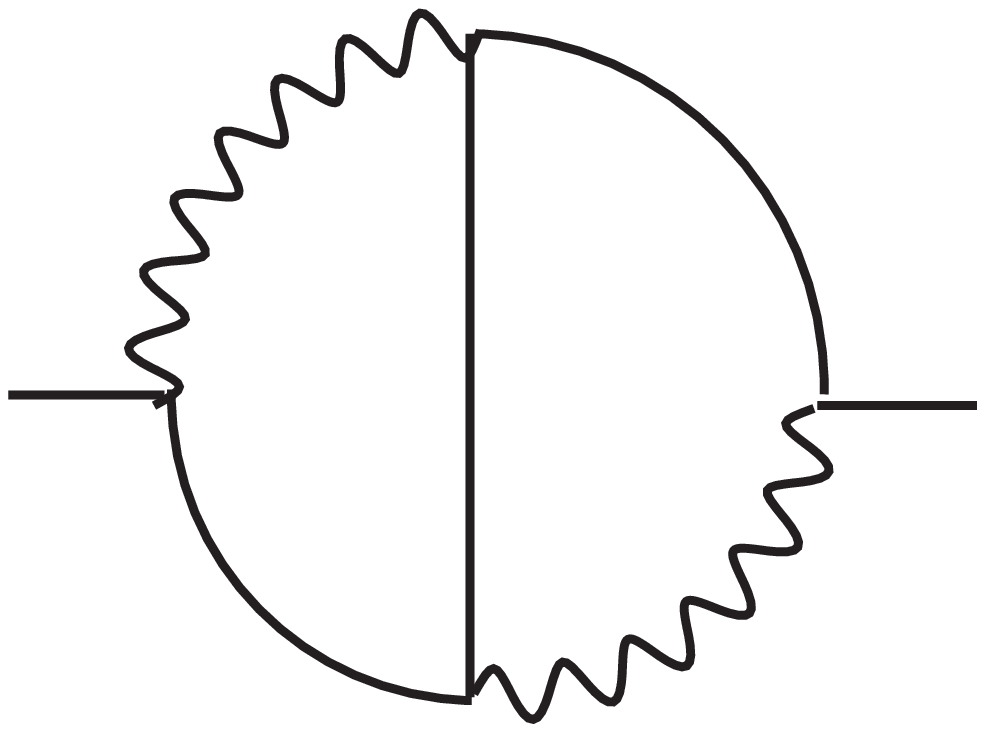,width=4cm}
\hspace*{1cm} 
\epsfig{file=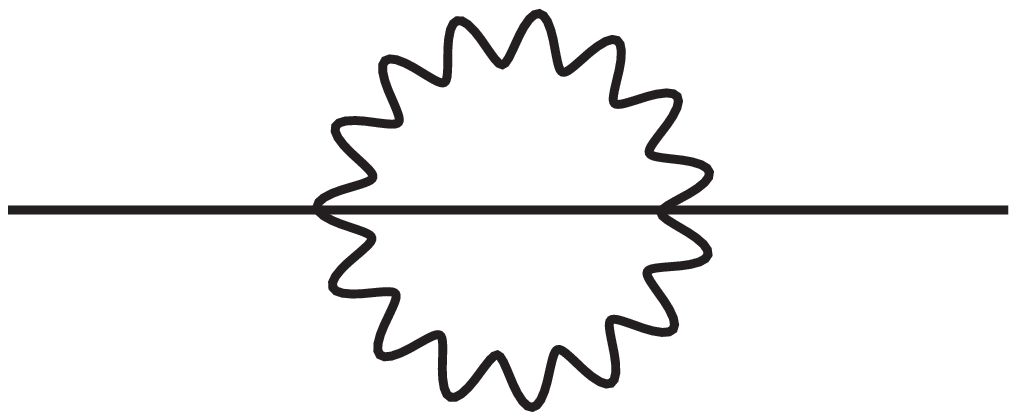,width=6cm}
  \end{center}
  \caption[On-shell self-energies; left: SE5l3m2 (example13.nb), right: SE3l1m (example14.nb)]{\label{f01101}
On-shell self-energies; left: SE5l3m2, right: SE3l1m}
\end{figure}

Mellin-Barnes representations can be also useful for solving on-shell
topologies. 
For on-shell self-energies, one may use the package {\ttfamily ON-SHELL2} \cite{Fleischer:1999tu,Kalmykov:web} written in FORM v.2.3  \cite{Vermaseren:1991??}.
In  {\ttfamily example13.nb} we show how to evaluate the self-energy {\tt SE5l3m2} shown in figure \ref{f01101}, which is in the notations of \cite{Fleischer:1999tu} the diagram {\ttfamily F01101}.
The MB-representation is a two-dimensional integral:
\begin{eqnarray}
{\rm F01101} &=&
-  \frac{1}{(2\pi i)^2} \frac{1}{\Gamma[1 - 2 \epsilon]}
  \int_{-i \infty}^{+i \infty} d z_1
  \int_{-i \infty}^{+i \infty} d z_2 \frac{\Gamma[-z_1] \Gamma[-z_2]
\Gamma[-\epsilon - z_1]
\Gamma[1 + 2 \epsilon + z_1]}
{ \Gamma[1 - 3 \epsilon - z_1]}
\nonumber \\
&& \times \frac{  
\Gamma[-\epsilon - z_1 - z_2]  
   \Gamma[-3 \epsilon - z_1 + z_2] \Gamma[1 - \epsilon + z_1 + z_2]
\Gamma[1 + \epsilon + z_1 + z_2]}
{  \Gamma[1 - \epsilon - z_1 + z_2] \Gamma[2 + \epsilon + z_1 + z_2]} .
\end{eqnarray}
In the example, the  agreement with the result of {\ttfamily On-Shell2} result is demonstrated.

A simpler case is {\tt SE3l1m} with one
massive and two massless
propagators, see figure \ref{f01101}. The result is simple:
\begin{equation}
{\rm SE3l1mOS} = -(m^2)^{(1 - 2 \epsilon)} \frac{\Gamma[3 - 4
\epsilon]
\Gamma[1 - \epsilon]^2 \Gamma[\epsilon] \Gamma[-1 + 2 \epsilon]}
{\Gamma[3 - 3 \epsilon] \Gamma[2 - 2 \epsilon]} .
\end{equation}
It can be expanded easily to any order in $\epsilon$. 
Here, in {\ttfamily example14.nb}, Barnes' second lemma has been used, which
happens not too often. 
Again, the agreement with the {\ttfamily On-Shell2}  result is presented.

\section{\label{sec:nonpl}Non-planar topologies}
\begin{figure}[thbp]
  \begin{center}
    \epsfig{file=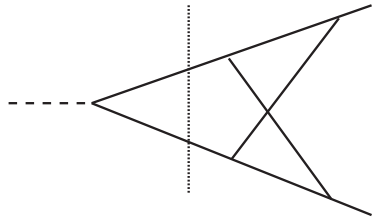,width=5cm}
  \end{center}
  \caption{\label{vnp}
Non-planar massless vertex}
\end{figure}

The loop-by-loop iterative procedure described in this paper seems to be
not the most efficient approach in the case of non-planar topologies.
It is known from \cite{Gonsalves:1983nq} that the massless non-planar vertex
is described by a two-dimensional Feynman parameter integral.
If we consider the loop-by-loop procedure for this case, 
we can divide the two-loop topology in figure \ref{vnp} into two parts (follow the vertical line). 
The hourglass
topology on the right-hand side, with two off-shell legs, gives a three-dimensional
MB-representation \cite{Smirnov:2006ry}, and adding the second part on the left-hand side
we end up with a four-dimensional integral. 
No matter how we arrange the momenta
flows in the diagram, it cannot become better. 
To get the minimal,
two-dimensional integral, another approach must be realized.
It is an open question to us
if the representation of non-planar diagrams can be automatized in a way like that for planar cases\footnote{The non-planar examples in the study \cite{Anastasiou:2005cb} do not go beyond our observations stated here.}.

\section{\label{sec:concl}Summary}
We have described the Mathematica package \ar{} for
the construction of MB-represen\-ta\-tions for planar 
Feynman integrals and gave in a tutorial part a variety of sample applications. 
Typically, the iterative loop-by-loop approach 
gives a possibility to construct MB-integrals of minimal dimension.
Usually Barnes' first and second lemmas help to get the minimal 
dimension of MB-integrals, independent both of the flow of momenta in 
diagrams and of the order of iterations. 
However, for more complicated 
kinematics, starting with five legs, the order of iterations and the choice of
momenta flows matters. 
As is shown there in the case of tadpoles, MB-representations
for massive topologies are not always the best way of evaluation.
For some topologies quite simple representations are found, however,
also multi-dimensional MB-integrals may arise from which it is hard 
to get stable,   accurate numerical results, not mentioning exact analytical results.

Constructing useful MB-representations for a given Feynman integral 
is a kind of an art.
As an example, let us mention
the QED master integral {\ttfamily B5l2m2} (a diagram with five lines, two of them being massive; notations are due to \cite{Czakon:2004wm,Czakon:2006pa}). 
This integral may be  obtained
by contracting directly two lines
in the massive Bhabha two-loop planar integral {\ttfamily B7l4m1} \cite{Smirnov:2001cm} (the so-called first planar master of massive QED).
In \cite{Czakon:2006pa} it was shown that, 
after expansion in $\epsilon$, 
the expression for {\tt B5l2m2}  consists of eleven integrals, one being four-dimensional.
This was compared to constructing {\tt B5l2m2} from the scratch, loop-by-loop.
Here, again after expansion in $\epsilon$, we are left with four integrals, all of them being three-dimensional or simpler. 
This can be checked easily by the reader using 
the \mb{} package 
and both the representation {\ttfamily B7l4m1} given in \cite{Smirnov:2001cm} 
(the contraction of two lines must be done there) 
and the representation {\tt B5l2m2}  given in \cite{Czakon:2006pa}. 
There is no simple relation between both representations,
due to our lack of knowledge on
more complicated relations between integrals of different dimensionality.
One should also mention that an independent check of the MB-representations always is strongly recommended.
By construction, neither \ar{} nor \mb{} perform rigorous proofs of their applicability. 

Certainly, the number of integration variables is of importance for the final 
evaluation of MB-integrals, both in a fully analytical form 
or using approximations in some kinematical limits. 
In many cases some package like {\tt XSUMMER} \cite{Moch:2001zr,Moch:2005uc}
can be used after deriving sums over residua.
This again might become non-effective if the number of the nested sums 
-- connected with the dimension of the MB-integrals -- is too large
or if the result is not in the class of functions covered by (e.g.) {\tt XSUMMER}.
Similar statements hold for the case of a fully numerical evaluation of MB-integrals.

To summarize, for many applications of present phenomenological or more theoretical interest
the package \ar{} solves an important  part of the complete calculational problem: 
the derivation of expressions for a large class of Feynman integrals, which may then be used for further study.

\section*{Acknowledgments}

We would like to thank Sven Moch and Bas Tausk for useful discussions.
The development of this package profited very much from a long cooperation with
Michal Czakon on NNLO corrections to Bhabha scattering in QED.
The present work is supported in part by the European Community's Marie-Curie Research Training Networks  MRTN-CT-2006-035505 `HEPTOOLS'
%
and MRTN-CT-2006-035482 `FLAVIAnet', 
and by  
Sonderforschungsbe\-reich/Trans\-regio 9--03 of Deutsche Forschungsgemeinschaft
`Computergest{\"u}tzte Theo\-re\-ti\-sche Teil\-chen\-phy\-sik'. 

\appendix
\section{\label{help}\ar{} functions list}
The basic functions of \ar{} are:
\begin{itemize}
\item
{\bf Fullintegral[\{numerator\},\{propagators\},\{internal momenta\}]} -- is the basic function for 
input Feynman integrals
\\
\item
{\bf invariants} -- is a list of invariants, e.g. {\bf invariants = \{p1*p1 $\to$ s\} }
\\
\item
{\bf IntPart[iteration]} -- prepares a subintegral for a given internal momentum by collecting the related numerator, propagators, integration momentum
\\
\item
{\bf Subloop[integral]} -- determines for the selected subintegral the $U$ and $F$ polynomials and an MB-representation
\\
\item
{\bf ARint[result,i\_]} -- displays the MB-representation number i for Feynman integrals
with numerators
\\
\item
{\bf Fauto[0]} -- allows user specified modifications of the $F$ polynomial {\bf fupc}
\\
\item
{\bf BarnesLemma[repr,1,Shifts\verb+->+True]}
-- function tries to apply Barnes' first lemma to a given 
MB-representation; when {\bf Shifts\verb+->+True} is set, \ar{} will try a simplifying shift of variables
\\ 
{\bf BarnesLemma[repr,2,Shifts\verb+->+True]}
-- function tries to apply Barnes' second lemma
\end{itemize}

\clearpage


\providecommand{\href}[2]{#2}


\end{document}